\documentclass[prd,superscriptaddress,amsfonts,amssymb,amsmath,showpacs,onecolumn]{revtex4}
\usepackage{bm}
\usepackage{amsfonts}
\usepackage{latexsym}
\usepackage[latin1]{inputenc}
\usepackage{graphicx}
\usepackage{amsmath}
\usepackage{palatino}
\usepackage{mathpazo}
\usepackage{textcomp}
\usepackage{float}
\usepackage{booktabs}
\usepackage{dcolumn}
\usepackage{hyperref}
\hypersetup{colorlinks,citecolor=blue}
\usepackage{amsmath}
\usepackage{xcolor}
\usepackage{mathtools}
\usepackage{orcidlink}
\usepackage[caption=false]{subfig}
\usepackage{commath}
\def\jnl@style{\it}
\def\aaref@jnl#1{{\jnl@style#1}}

\def\aaref@jnl#1{{\jnl@style#1}}

\def\aj{\aaref@jnl{AJ}}                   % Astronomical Journal
\def\apj{\aaref@jnl{ApJ}}                 % Astrophysical Journal
\def\apjl{\aaref@jnl{ApJ}}                % Astrophysical Journal, Letters
\def\apjs{\aaref@jnl{ApJS}}               % Astrophysical Journal, Supplement
\def\apss{\aaref@jnl{Ap\&SS}}             % Astrophysics and Space Science
\def\aap{\aaref@jnl{A\&A}}                % Astronomy and Astrophysics
\def\aapr{\aaref@jnl{A\&A~Rev.}}          % Astronomy and Astrophysics Reviews
\def\aaps{\aaref@jnl{A\&AS}}              % Astronomy and Astrophysics, Supplement
\def\mnras{\aaref@jnl{Mon.~Not.~Roy.~Astron.~Soc.}}             % Monthly Notices of the RAS
\def\prd{\aaref@jnl{Phys.~Rev.~D}}        % Physical Review D
\def\prc{\aaref@jnl{Phys.~Rev.~C}}  % Physical Review C
\def\prl{\aaref@jnl{Phys.~Rev.~Lett.}}    % Physical Review Letters
\def\qjras{\aaref@jnl{QJRAS}}             % Quarterly Journal of the RAS
\def\skytel{\aaref@jnl{S\&T}}             % Sky and Telescope
\def\ssr{\aaref@jnl{Space~Sci.~Rev.}}     % Space Science Reviews
\def\zap{\aaref@jnl{ZAp}}                 % Zeitschrift fuer Astrophysik
\def\nat{\aaref@jnl{Nature}}              % Nature
\def\aplett{\aaref@jnl{Astrophys.~Lett.}} % Astrophysics Letters
\def\apspr{\aaref@jnl{Astrophys.~Space~Phys.~Res.}} % Astrophysics Space Physics Research
\def\physrep{\aaref@jnl{Phys.~Rep.}}      % Physics Reports
\def\physscr{\aaref@jnl{Phys.~Scr}}       % Physica Scripta
\def\commat{\aaref@jnl{Comm.~Math.~Phys.}}              % Communications in Mathematical Physics
\def\science{\aaref@jnl{Science}}               % Science
\def\cqg{\aaref@jnl{Classical Quant.~Grav.}}            % Classical and Quantum Gravity
\def\jpcs{\aaref@jnl{JPCS}}                                     % Journal of Physics Conference Series
\def\ijmpd{\aaref@jnl{Int.~J.~Mod.~Phys.~D}}                    % International Journal of Modern Physics D
\def\grg{\aaref@jnl{Gen.~Relat.~Gravit.}}               % General Relativity and Gravitation
\def\rpp{\aaref@jnl{Rep.~Prog.~Phys.}}          % Reports on Progress in Physics
\def\npa{\aaref@jnl{Nucl.~Phys.~A}}        % Nuclear Physics A
\def\lrr{\aaref@jnl{Living Rev.~Rel.}}                   % Living reviews in relativity
\def\jcap{\aaref@jnl{J.~Cosmology Astropart.~Phys.}}    % Journal of cosmology and astroparticle physics
\def\rmp{\aaref@jnl{Rev.~Mod.~Phys.}}   %Reviews of modern physics

\allowdisplaybreaks[1]
\begin{document}
%\color{red}
\color{black}       %% For one column

\title{Late-Time Viscous Cosmology in $f(R,T)$ Gravity}

\author{Simran Arora\orcidlink{0000-0003-0326-8945}}
\email{dawrasimran27@gmail.com}
\affiliation{Department of Mathematics, Birla Institute of Technology and
Science-Pilani,\\ Hyderabad Campus, Hyderabad-500078, India.}
\author{Snehasish Bhattacharjee\orcidlink{0000-0002-7350-7043}}
\email{snehasish.bhattacharjee.666@gmail.com}
\affiliation{Department of Astronomy, Osmania University, Hyderabad-500007, India.}
\author{P.K. Sahoo\orcidlink{0000-0003-2130-8832}}
\email{pksahoo@hyderabad.bits-pilani.ac.in}
\affiliation{Department of Mathematics, Birla Institute of Technology and
Science-Pilani,\\ Hyderabad Campus, Hyderabad-500078, India.}
%%%%%%%%%%%%%%%%%%%%%%%%%%%%%%%%%%%%  DATE  %%%%%%%%%%%%%%%%%%%%%%%%%%%%%%%%%%%%
\date{\today}

\begin{abstract}
\textbf{Abstract:} The article communicates an alternative route to suffice the late-time acceleration considering a bulk viscous fluid with viscosity coefficient $\zeta =\zeta _{0}+ \zeta _{1} H + \zeta _{2} H^{2}$, where $\zeta _{0}, \zeta _{1}, \zeta _{2}$ are constants in the framework of $f(R,T)$ modified gravity. We presume the $f(R,T)$ functional form to be $f=R+2\alpha T$ where $\alpha$ is a constant. We then solve the field equations for the Hubble Parameter and study the cosmological dynamics of kinematic variables such as deceleration, jerk, snap and lerk parameters as a function of cosmic time. We observe the deceleration parameter to be highly sensitive to $\alpha$ and undergoes a signature flipping at around $t\sim 10$ Gyrs for $\alpha=-0.179$ which is favored by observations. The EoS parameter for our model assumes values close to $-1$ at $t_{0}=13.7$Gyrs which is in remarkable agreement with the latest Planck measurements. Next, we study the evolution of energy conditions and find that our model violate the Strong Energy Condition in order to explain the late-time cosmic acceleration. To understand the nature of dark energy mimicked by the bulk viscous baryonic fluid, we perform some geometrical diagnostics like the  $\{r,s\}$ and $\{r,q\}$ plane. We found the model to mimic the nature of a Chaplygin gas type dark energy model at early times while a Quintessence type in distant future. Finally, we  study the violation of continuity equation for our model and show that in order to explain the cosmic acceleration at the present epoch, energy-momentum must violate.

\textbf{Keywords:} $f(R,T)$ gravity; Bulk Viscosity; Energy Conditions; Statefinder parameters
\end{abstract}

\keywords{$f(R,T)$ gravity; Equation of State; Bulk Viscosity; Energy Conditions; Statefinder parameters}
\pacs{95.36.+x, 04.50.kd, 98.80.Jk.}

\maketitle

\section{Introduction}\label{I}

Multiple observations confirm that at present time, the universe is experiencing a phase of accelerated expansion \cite{Riess98,Per99}. The enigmatic entity called ``Dark Energy" with an EoS parameter $\omega\simeq-1$ is presumed to be the culprit for such an accelerated expansion. ``Dark Energy" possess negative pressure and thence create antigravity effect which permeate in all of spacetime, owing to which cosmic structures separated by Mpcs manages to overcome their mutual gravitational forces and fly apart from each other. Nonetheless, no conclusive evidence have yet emerged to solidify the ominous presence of ``Dark Energy". As a consequence, many alternate models have emerged to explain this conundrum \cite{Ratra88,cal88,Buc00,Arm01,Tom01,Mil03,hunt10,Eas11,Rad12,Rad13,Pan17,Pan19}.\\
In majority of the cosmological models, the cosmic fluid is presumed to be devoid of any viscous (shear and bulk) which greatly simplifies the field equations. Such simplifications may seem plausible under most circumstances but not at all times. For instance, when fluid motion near solid boundaries is considered \cite{Brevik11}. In the context of cosmology, the cosmic fluid to a large extent is spatially isotropic and therefore the shear viscosity plays no role in cosmic dynamics. Having said that, the bulk viscosity could play a very important role in governing the cosmic evolution by modifying the background dynamics \cite{Almada20,singh}. Some of the earliest works in bulk viscous cosmology dates back to 1970's \cite{Misner68,Israel,Murphy,Belinskii}. Around 1980's, pioneering studies reported the possibility of inflation being driven by bulk viscous fluids \cite{Waga,Barrow86,Barrow88}. The phenomenon that ``Dark Energy" could be an effect of bulk viscosity in the cosmic medium was reported in \cite{singh48}. Bulk viscous fluids have also been reported to be promising candidates for ``Dark Matter" \cite{singh49}, ``Dark Energy" \cite{Cataldo05,Brevik05,Setare10,Gagnon11} and unified scenarios \cite{Li09,Hipolito09,Hipolito10,Montiel11,Fabris11,Velten11}. Other interesting studies can be found in \cite{Fabris06,Kremer12,Avelino13,singh,Atreya18,Valentino19} (aslo see \cite{pan55} for a recent review on bulk viscous cosmology).\\
In this article we investigate the possibility that an accelerated expansion ought to be possible owing to the presence of a bulk viscous baryonic fluid. We therefore turn our attention to modified gravity theories (MGT) which refute the existence of the ``Dark Energy" and ``Dark Matter" by presuming them to be purely geometrical in nature. MGTs are simple geometrical extensions of General Relativity. The action here is altered by substituting the Ricci scalar $R$ with other curvature invariants such as Torsion scalar $\mathcal{T}$, Gauss-Bonnet scalar $\mathcal{G}$, non-metricity  $\mathcal{Q}$ etc.   \\  
In this work we shall work with $f(R,T)$ gravity theory in which the Ricci scalar $R$ is replaced with a suitable functional form of $R$ and trace of energy momentum tensor $T$ \cite{harko} and therefore is a straightforward conjecture to $f(R)$ gravity (see \cite{cap02,cap11}). $f(R, T)$ gravity have proved to be successful in numerous cosmological sectors such as dark matter \cite{in22} dark energy \cite{in21}, massive pulsars \cite{in23,santos19}, super-Chandrasekhar white dwarfs \cite{in25}, wormholes \cite{Aziz13,moraes2017,moraes17,Yousaf2017,Sahoo2018,Sahoo18,moraes18,Elizalde18,in26,moraes19,moraes/19}, gravitational waves \cite{Alves,in36,Bhatti20}, bouncing cosmology \cite{bounce,bounce2}, baryogenesis \cite{baryo,baryo2}, Big-Bang nucleosynthesis \cite{bang} and in varying speed of light scenarios \cite{physical}.\\
The article is methodized as follows: In Section \ref{II} we provide a summary of $f(R,T)$ gravity and solve the field equations assuming a bulk viscous fluid. In Section \ref{III} we study the temporal evolution of kinematic variables. In Section \ref{IV} we investigate the evolution of energy-density, effective pressure and EoS parameter. In Section \ref{V} we study the growth of various energy conditions. Whereas in Section \ref{VI} we perform some geometrical diagnostics for our model. In Section \ref{VII} we study the violation of energy-momentum and in Section \ref{VIII} we conclude with our results.

\section{Overview of $f(R,T)$ Gravity}\label{II}
The action in $f(R, T)$ gravity is given as \citep{harko}

\begin{equation}  \label{e1}
S=\frac{1}{2}\int d^{4}x\sqrt{-g}(f(R,T)+ 2L_{m}),
\end{equation}
here $g$ is the metric determinant and $L_{m}$ the matter Lagrangian.

Variation of the above action with respect to the metric tensor gives

\begin{multline}  \label{e2}
f_{R}(R,T)R_{\mu \nu}-\frac{1}{2} f(R,T)g_{\mu \nu}+(g_{\mu
\nu}\square-\nabla_{\mu} \nabla_{\nu}) f_{R}(R,T)\\ = T_{\mu
\nu}-f_{T}(R,T)T_{\mu \nu}-f_{T}(R,T)\Theta_{\mu \nu},
\end{multline}
where, $\nabla _{\mu }$ and $\nabla _{\nu }$ represents the covariant
derivative and $\Theta _{\mu \nu }$ is defined by 
\begin{equation}
\Theta _{\mu \nu }\equiv g^{\alpha \beta }\frac{\delta T_{\alpha \beta }}{%
\delta g^{\mu \nu }}.  \label{e3}
\end{equation}

We shall consider a flat spacetime geometry 
\begin{equation}  \label{e4}
ds^{2}= dt^{2}-a^{2}(t)[dr^{2}+r^{2}d\theta^{2}+r^{2}\sin^{2}\theta
d\phi^{2}]
\end{equation}
where $a(t)$ is the cosmic scale factor. 

the components of four-velocity $u^{\mu}$ are $u^{\mu}=(1,0)$ in comoving
coordinates. Assume that the cosmic fluid possesses a bulk viscosity $\zeta$%. We have the energy-momentum tensor for a viscous fluid as follows 
\begin{equation}  \label{e5}
T_{\mu \nu}= \rho u_{\mu}u_{\nu}-\overline{p} h_{\mu \nu}.
\end{equation}
where $h_{\mu \nu}= g_{\mu \nu}+ u_{\mu}u_{\nu}$ and $\overline{p}= p-3\zeta
H$ is the effective pressure.

If we choose the Lagrangian density as $L_{m}= -\overline{p}$ then the
tensor $\Theta_{\mu \nu}$ becomes 
\begin{equation}  \label{e6}
\Theta_{\mu \nu}= -2T_{\mu \nu}-\overline{p} g_{\mu \nu}.
\end{equation}
using (\ref{e5}) and (\ref{e6}), the field equation for the bulk viscous
fluid become 
\begin{equation}  \label{e7}
R_{\mu \nu}-\frac{1}{2}Rg_{\mu \nu}= T_{\mu \nu}+ 2f^{\prime }(T)T_{\mu
\nu}+(2 \overline{p}f^{\prime }(T)+f(T)) g_{\mu\nu}.
\end{equation}

For the particular choice of the function $f(T)=\alpha T$, where $\alpha $
is a constant, we get field equations as 
\begin{equation}
3H^{2}=\rho +2\alpha (\rho +\overline{p})+\alpha T,  \label{e8}
\end{equation}%
\begin{equation}
2\dot{H}+3H^{2}=-\overline{p}+\alpha T,  \label{e9}
\end{equation}%
where $T=\rho -3\overline{p}$. From Eqs. (\ref{e8}) and (\ref{e9}), we have 
\begin{equation}
\ 2\dot{H}+(1+2\alpha )(p+\rho )-3(1+2\alpha )\zeta H=0.  \label{e10}
\end{equation}

\subsection{Bulk Viscous Solutions}

Note that the Eqs. (\ref{e8}) and (\ref{e9}) contain four unknown
parameters viz. $\rho, p, \zeta  \ \& H$. To get an exact solution, two
more supplementary equations are necessary. Bearing that in mind, we consider the following relationship between pressure and density  
\begin{equation}
p=(\gamma -1)\rho,
\label{e11}
\end{equation}
where $0<\gamma <1$ is a constant.\\
Additionally, we assume the bulk viscosity coefficient $\zeta$ as
\begin{equation}
\zeta =\zeta _{0}+ \zeta _{1} H + \zeta _{2} H^{2}. 
\label{e12}
\end{equation}
 $\zeta _{0}, \zeta _{1}, \zeta _{2}$ are constants. 

Employing Eqs. (\ref{e8}), (\ref{e11}), (\ref{e12}), the
expression of energy density reads
\begin{equation}
\rho = \frac{3 ((1-\alpha \zeta_{1})H^{2}- \alpha \zeta_{0}H- \alpha \zeta_{2} H^{3})}{(1+4\alpha-\alpha\gamma)}.
\label{e13}
\end{equation}
Furthermore, from Eqs. (\ref{e10}), (\ref{e11}), (\ref{e13}), we arrive at the following equation

\begin{equation}\label{e14} 
2 \dot{H} - 3(1+2\alpha)\left[ \frac{\zeta_{0}(1+4\alpha)}{(1+4\alpha-\alpha\gamma)}\right]  H +3(1+2\alpha)\left[ \frac{\gamma-(1+4\alpha)\zeta_{1}}{1+4\alpha-\alpha\gamma}\right] H^{2}-3(1+2\alpha)\left[ \frac{\zeta_{2}(1+4\alpha)}{1+4\alpha-\alpha\gamma)}\right] H^{3}=0.
\end{equation}%

Owing to the high non-linearity of the above equation, we restrict ourselves to terms with leading orders in $H$ which yields 
\begin{equation}
2\dot{H}+ \left[ \frac{3(1+2\alpha)(1+4\alpha)\zeta_{2}}{(1+4\alpha-\alpha\gamma)}\right] H^{3}=0. \label{e15}
\end{equation}

Solving for the Hubble parameter $H$ yields
parameter $H$ as, 
\begin{equation}
H=\frac{1}{\sqrt{k_{1}t+c}}, \label{e16}
\end{equation}%
where $k_{1}=  \frac{3(1+2\alpha )\zeta_{2}(1+4\alpha )}{1+4\alpha -\alpha
\gamma }$, with $c$ being a constant of integration.

From the relation $H=\frac{\dot{a}}{a}$, we obtain the scale factor as 
\begin{equation}
a(t)= c_{1} e^{(2/k_{1})\sqrt{k_{1}t+c}}, \label{e17}
\end{equation}
where $c_{1}$ is an integrating constant.

The deceleration parameter $q$ upon using $q=-1-\frac{\dot{H}}{H^{2}}$ reads, 
\begin{equation}
q= -1+ \frac{k_{1}}{2 \sqrt{k_{1}t+c}}. \label{e18}
 \end{equation}
\section{Kinematic variables}\label{III}

\begin{figure}[H]
\centering
\includegraphics[scale =0.6]{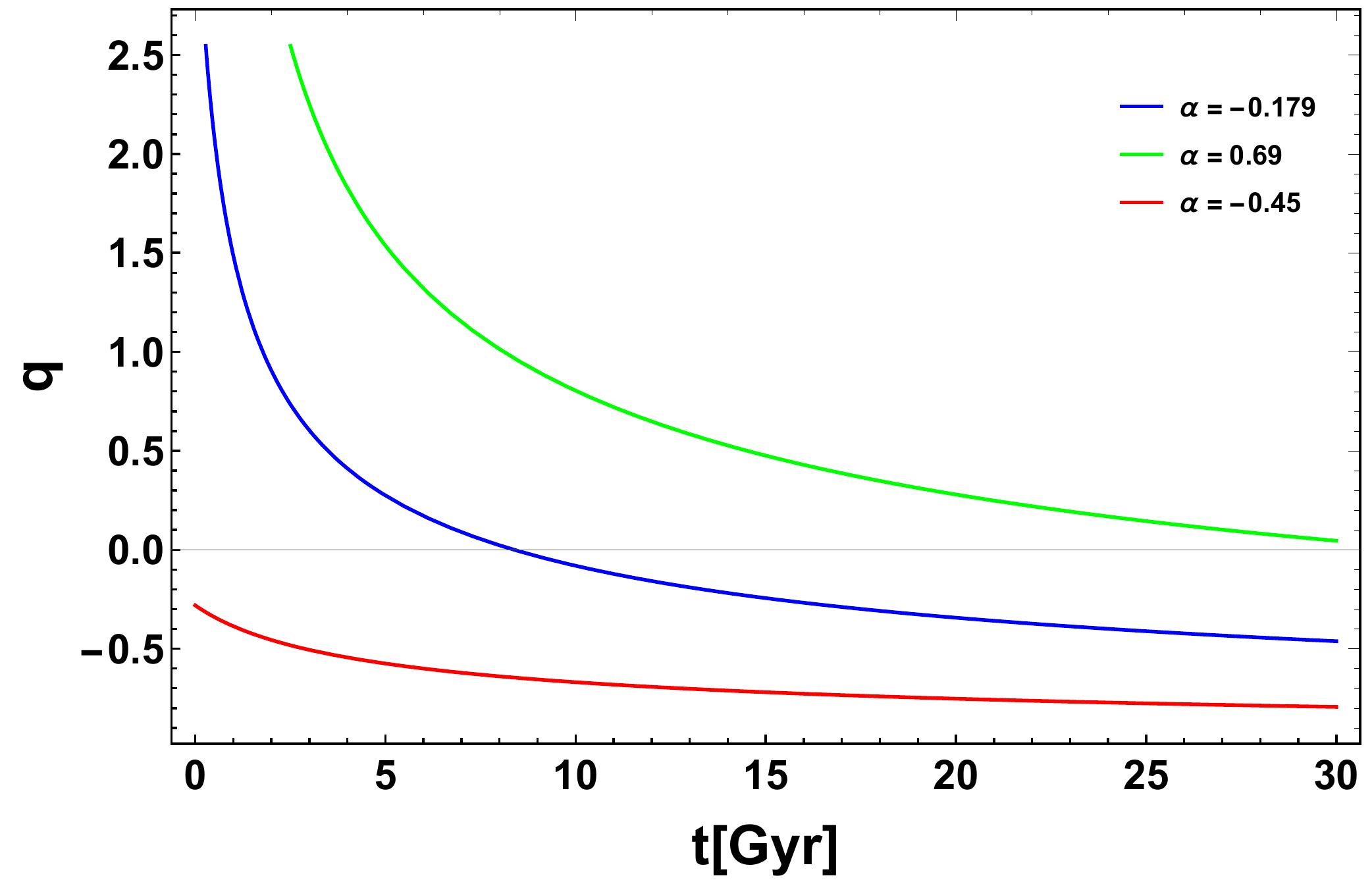}
\caption{Time variation of deceleration parameter.}
\label{fig1}
\end{figure}

The evolution of the deceleration parameter ($q$) is shown in Fig. \ref{fig1}. We observe $q$ to be highly sensitive to the model parameter $\alpha$ and undergoes a change in signature at around $t\sim 10$Gyrs for $\alpha=-0.179$ which is favored by observations. Interestingly, for other values of $\alpha$, the signature flipping is not observed. For instance, when $\alpha = -0.45$, our model represents an eternal acceleration whereas for $\alpha=0.69$, $q>0$ and therefore imply an decelerating universe. Both of these scenarios are highly incompatible with observations. \\\\

\begin{center}
\begin{tabular}{ |p{3cm}||p{3cm}|p{3cm}|p{3cm}|  }
 \hline
 \multicolumn{4}{|c|}{Behavior of parameters at t=13.7} \\
 \hline
 $\alpha$ & $k_{1}$ & $q$ & c\\
 \hline
 $\alpha= -0.45$ & $k_{1}>0$ & $q<0$ & $c>0$ or $c<0$ \\
 $\alpha= 0.69$ & $k_{1}>0$ & $q>0$ & $c>0$ or $c<0$ \\
 $\alpha= -0.179$ & $k_{1}>0$ & q shows transition & $c>0$ or $c<0$ \\ 
 $\alpha <-0.5$ & $k_{1}<0$ & not available &  $c>0$ or $c<0$\\
$\alpha=-0.5$ & $k_{1}=0$ & $q=-1$ &  $c>0$ or $c<0$\\
 
 \hline
\end{tabular}\label{tab1}
\end{center}

The Taylor's expansion of scale factor to get the higher order derivatives of deceleration parameter such as jerk ($j$), snap ($k$) and lerk ($l$) are useful in understanding the dynamics of the universe and are defined as follows \cite{Visser,Capozziello,Mandal/2020}

\begin{equation}\label{j1}
j(t)= \frac{1}{a} \frac{d^{3}a}{dt^{3}}\left[ \frac{1}{a} \frac{da}{dt}\right] ^{-3},
\end{equation}

\begin{equation}\label{s1}
s(t)= \frac{1}{a} \frac{d^{4}a}{dt^{4}}\left[ \frac{1}{a} \frac{da}{dt}\right] ^{-4},
\end{equation}

\begin{equation}\label{l1}
l(t)= \frac{1}{a} \frac{d^{5}a}{dt^{5}}\left[ \frac{1}{a} \frac{da}{dt}\right] ^{-5}.
\end{equation}

Employing \eqref{e18} in \eqref{j1}, \eqref{s1} \& \eqref{l1}, the expressions of $j(t)$, $s(t)$ \& $l(t)$ reads respectively as
\begin{equation}
\displaystyle j(t)= -  \frac{k_{1} \left(6 \sqrt{c+k_{1} t}-3 k_{1}-4 t\right)-4 c}{4 (c+k_{1} t)}.
\end{equation}

\begin{equation}
\displaystyle s(t)=  \frac{k_{1} \left(6 k_{1} \left(5 \sqrt{c+k_{1} t}-4 t\right)+8 t \sqrt{c+k_{1} t}-15 k_{1}^2\right)+8 c \left(\sqrt{c+k_{1} t}-3 k_{1}\right)}{8 (c+k_{1} t)^{3/2}} . 
\end{equation}

\begin{equation}
\displaystyle l(t)=  \frac{16 c^2+k_{1}^2 \left(-30 k_{1}\left(7 \sqrt{c+k_{1}t}-6 t\right)+16 t \left(t-5 \sqrt{c+k_{1}t}\right)+105 k_{1}^2\right)+4 c k_{1} \left(-20 \sqrt{c+k_{1} t}+45 k_{1}+8 t\right)}{16 (c+k_{1} t)^2} .
\end{equation}

where $k_{1}= -\frac{3 (2 \alpha +1) (4 \alpha +1) \zeta_{2}}{\alpha  (-\gamma )+4 \alpha +1}$.
\begin{figure}[H]
\centering
\includegraphics[scale =0.5]{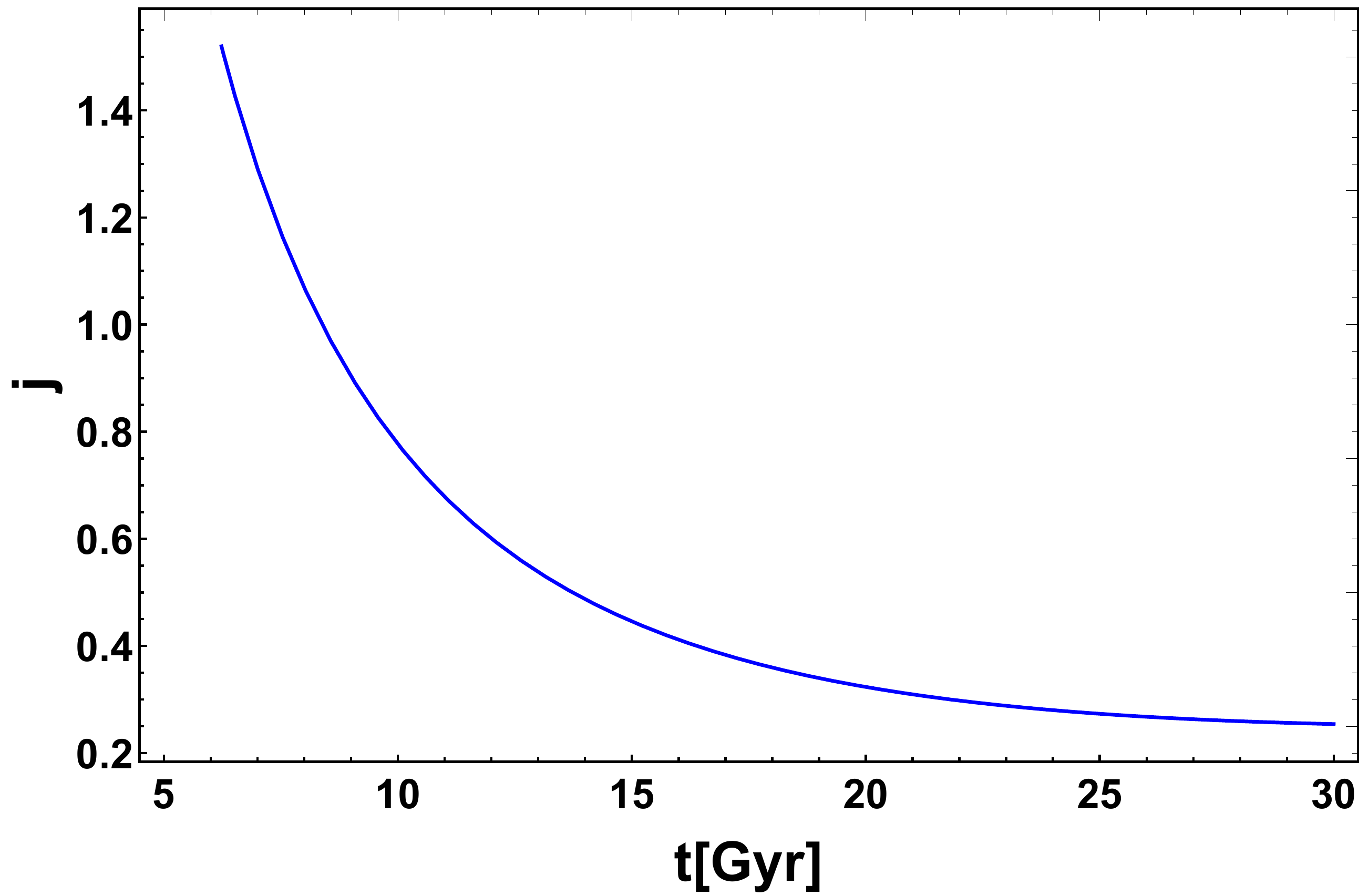}
\caption{Time variation of jerk parameter
with $\alpha=-0.179$, $\gamma=0.01$, $\zeta_{2}=-18.4$, c=14.9.}
\label{fig2}
\end{figure}

\begin{figure}[H]
\centering
\includegraphics[scale =0.5]{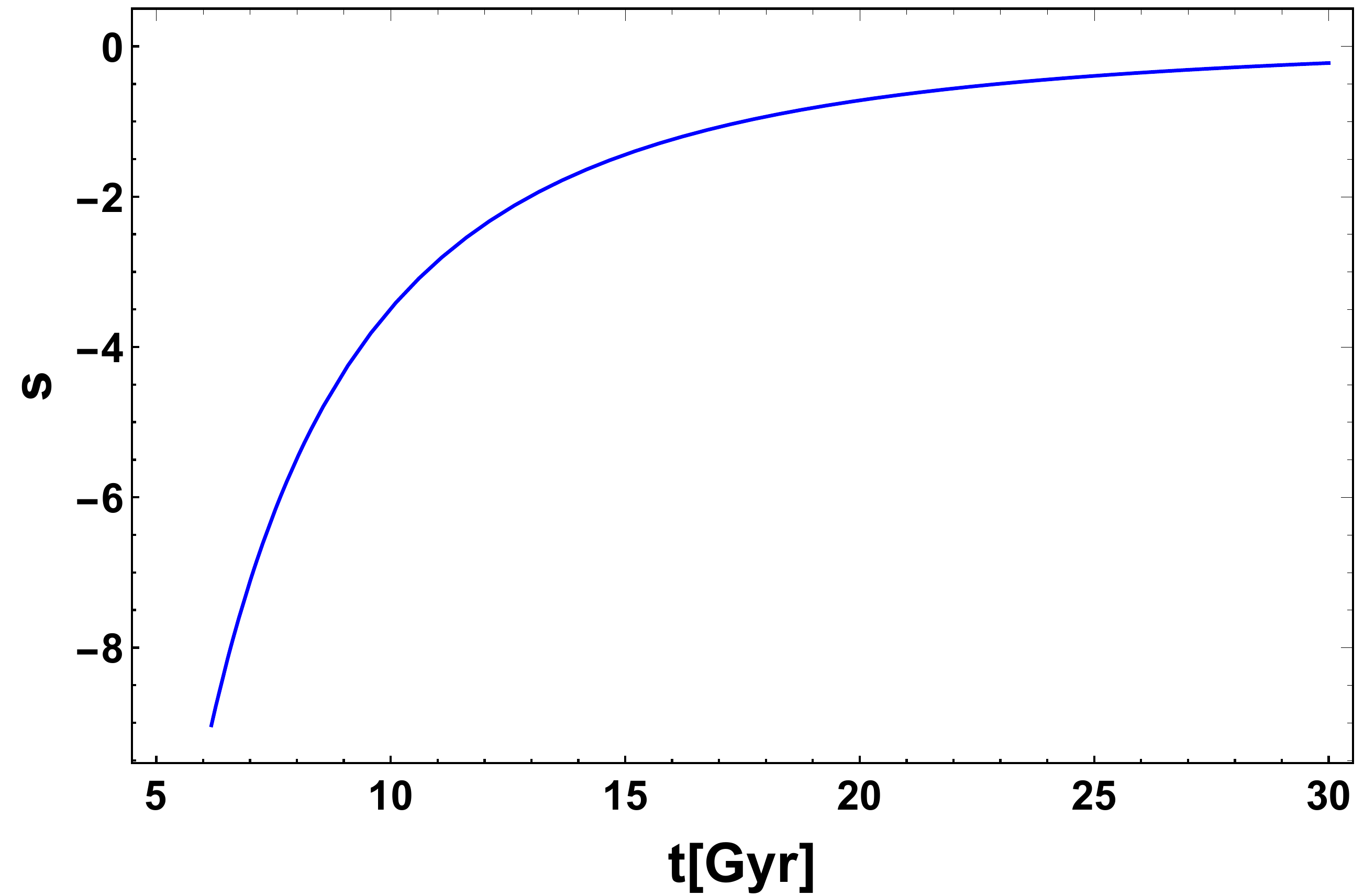}
\caption{Time variation of snap parameter
with $\alpha=-0.179$, $\gamma=0.01$, $\zeta_{2}=-18.4$, c=14.9.}
\label{fig3}
\end{figure}

 \begin{figure}[H]
\centering
\includegraphics[scale =0.5]{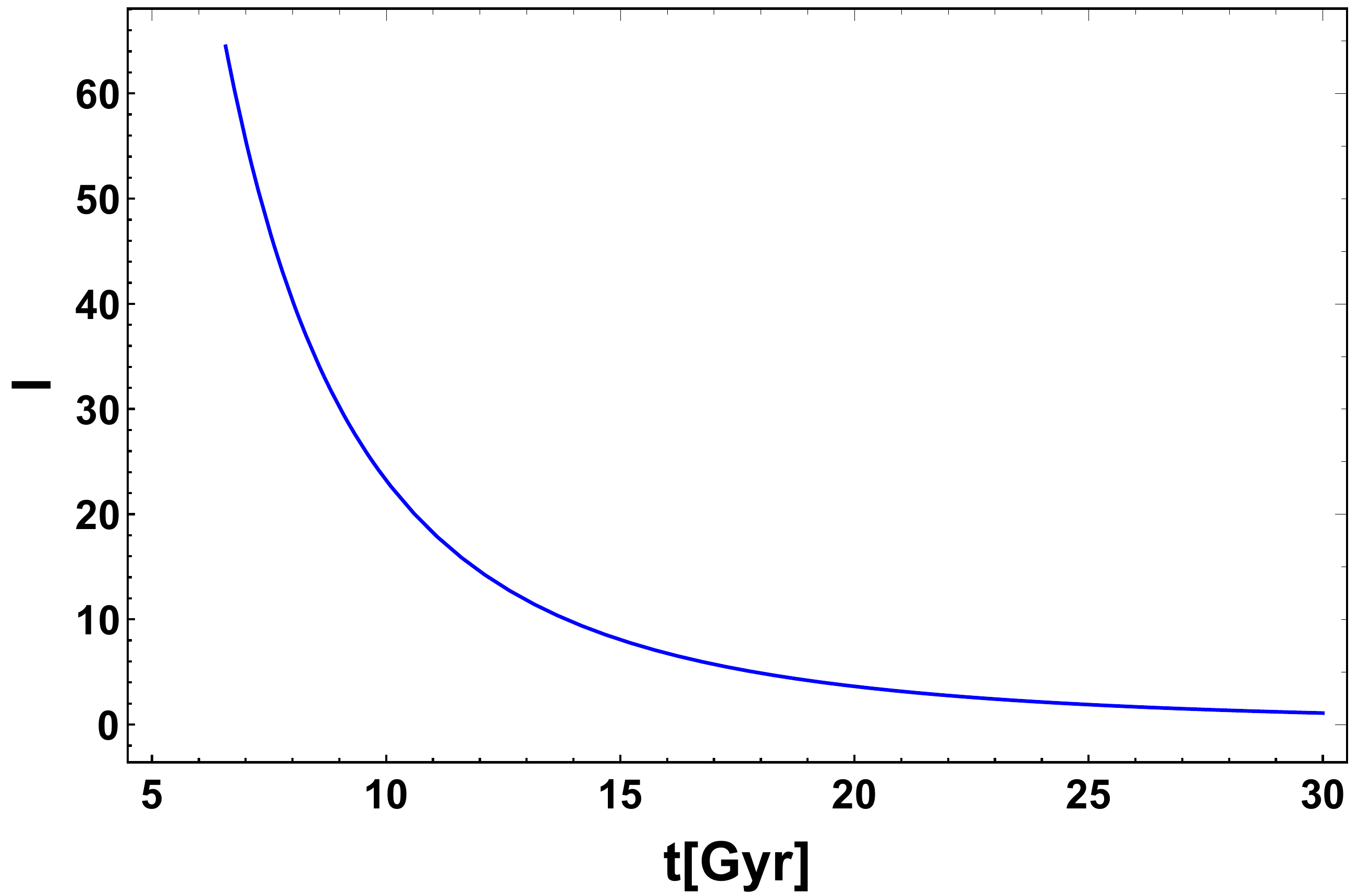}
\caption{Time variation of lerk parameter
with $\alpha=-0.179$, $\gamma=0.01$, $\zeta_{2}=-18.4$, c=14.9.}
\label{fig4}
\end{figure}
The rate of change of deceleration parameter is represented by the jerk parameter and therefore is a useful parameter to understand the future of the universe. These higher derivatives are also useful in understanding the emergence of sudden future singularities \cite{16}.\\
In Fig. \ref{fig2}, we show the temporal evolution of jerk parameter where the positivity of jerk parameter ensures an accelerated expansion. In Fig. \ref{fig3} the snap parameter is observed to undergo a signature flipping at around $t\sim8$ Gyr. In Fig. \ref{fig4}, we observe a positive lerk parameter for the entire cosmic aeon much like the jerk parameter. Interestingly, both the jerk and lerk are decreasing function of time whereas the snap is clearly an increasing function. We note that the magnitudes and behaviors of all these parameters strongly suggest an accelerating universe at the present epoch (i.e., $t=13.7$ Gyrs).

\section{The EoS Parameter}\label{IV}

In this section we shall explore the physical evolution of energy density, effective pressure and The equation of state(EoS) parameter. 

Substituting \eqref{e16} into \eqref{e13}, the expression of density $\rho$ reads

\begin{equation}
\displaystyle \rho=  \frac{-\frac{\alpha  \zeta_{0}}{\sqrt{c-\frac{3 (2 \alpha +1) (4 \alpha +1) \zeta_{2} t}{\alpha  (-\gamma )+4 \alpha +1}}}+\frac{3 (1-\alpha  \zeta_{1})}{c-\frac{3 (2 \alpha +1) (4 \alpha +1) \zeta_{2} t}{\alpha  (-\gamma )+4 \alpha +1}}-\frac{\alpha  \zeta_{2}}{\left(c-\frac{3 (2 \alpha +1) (4 \alpha +1) \zeta_{2} t}{\alpha  (-\gamma )+4 \alpha +1}\right)^{3/2}}}{(1+4 \alpha -\alpha \gamma)} .
\end{equation}
From the relation $\overline{p}= p-3\zeta
H$, the expression of effective pressure $\overline{p}$ reads
\begin{multline}
\displaystyle \overline{p}= \frac{(\gamma -1) \left(-\frac{\alpha  \zeta_{0}}{\sqrt{c-\frac{3 (2 \alpha +1) (4 \alpha +1) \zeta_{2} t}{\alpha  (-\gamma )+4 \alpha +1}}}+\frac{3 (1-\alpha  \zeta_{1})}{c-\frac{3 (2 \alpha +1) (4 \alpha +1) \zeta_{2} t}{\alpha  (-\gamma )+4 \alpha +1}}-\frac{\alpha  \zeta_{2}}{\left(c-\frac{3 (2 \alpha +1) (4 \alpha +1) \zeta_{2} t}{\alpha  (-\gamma )+4 \alpha +1}\right)^{3/2}}\right)}{\alpha  (-\gamma )+4 \alpha +1}-\frac{3 \zeta_{0}}{\sqrt{c-\frac{3 (2 \alpha +1) (4 \alpha +1) \zeta_{2} t}{\alpha  (-\gamma )+4 \alpha +1}}}\\
 -\frac{3 \zeta_{1}}{c-\frac{3 (2 \alpha +1) (4 \alpha +1) \zeta_{2} t}{\alpha  (-\gamma )+4 \alpha +1}}-\frac{3 \zeta_{2}}{\left(c-\frac{3 (2 \alpha +1) (4 \alpha +1) \zeta_{2} t}{\alpha  (-\gamma )+4 \alpha +1}\right)^{3/2}}. 
\end{multline}

Finally, the EoS parameter $\omega = \frac{\overline{p}}{\rho}$ reads.
\begin{equation}
 \omega= -\frac{\splitfrac{c \zeta_{0} (\alpha  (\gamma -4)-1) (\alpha  (2 \gamma -11)-3)-3 (\alpha  (\gamma -4)-1) (3 \alpha  \zeta_{1}-\gamma +\zeta_{1}+1) \sqrt{c+\frac{3 \left(8 \alpha ^2+6 \alpha +1\right) \zeta_{2} t}{\alpha  (\gamma -4)-1}}}{+\zeta_{2} (2 \alpha  \gamma -11 \alpha -3) (\alpha  (\gamma +6 (4 \alpha +3) \zeta_{0} t-4)+3 \zeta_{0} t-1)}}{\splitfrac{\alpha  c \zeta_{0} (\alpha  (\gamma -4)-1)+3 \alpha  (\zeta_{1} (\alpha  (\gamma -4)-1)-\gamma +4) \sqrt{c+\frac{3 \left(8 \alpha ^2+6 \alpha +1\right) \zeta_{2} t}{\alpha  (\gamma -4)-1}}+3 \sqrt{c+\frac{3 \left(8 \alpha ^2+6 \alpha +1\right) \zeta_{2} t}{\alpha  (\gamma -4)-1}}}{+\alpha  \zeta_{2} (\alpha  (\gamma +6 (4 \alpha +3) \zeta_{0} t-4)+3 \zeta_{0} t-1)}}.
\end{equation}

\begin{figure}[H]
\centering
\includegraphics[scale =0.5]{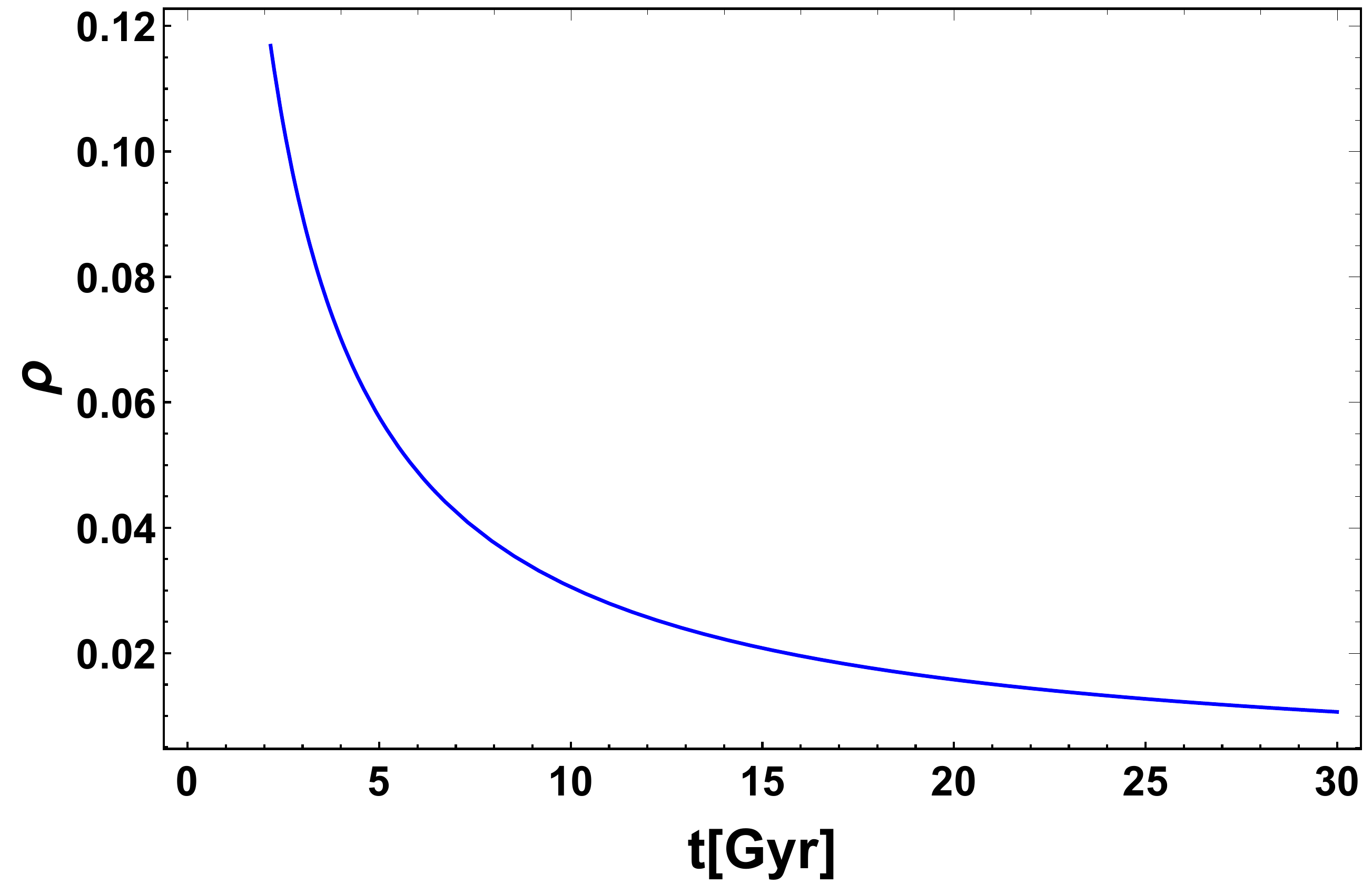}
\caption{Evolution of density 
with $\alpha=-0.179$, $\gamma=0.01$, $\zeta_{0}=-0.0076$, $\zeta_{1}= 0.75$, $\zeta_{2}=-18.4$, $c=14.9$.}
\label{fig5}
\end{figure}

\begin{figure}[H]
\centering
\includegraphics[scale =0.5]{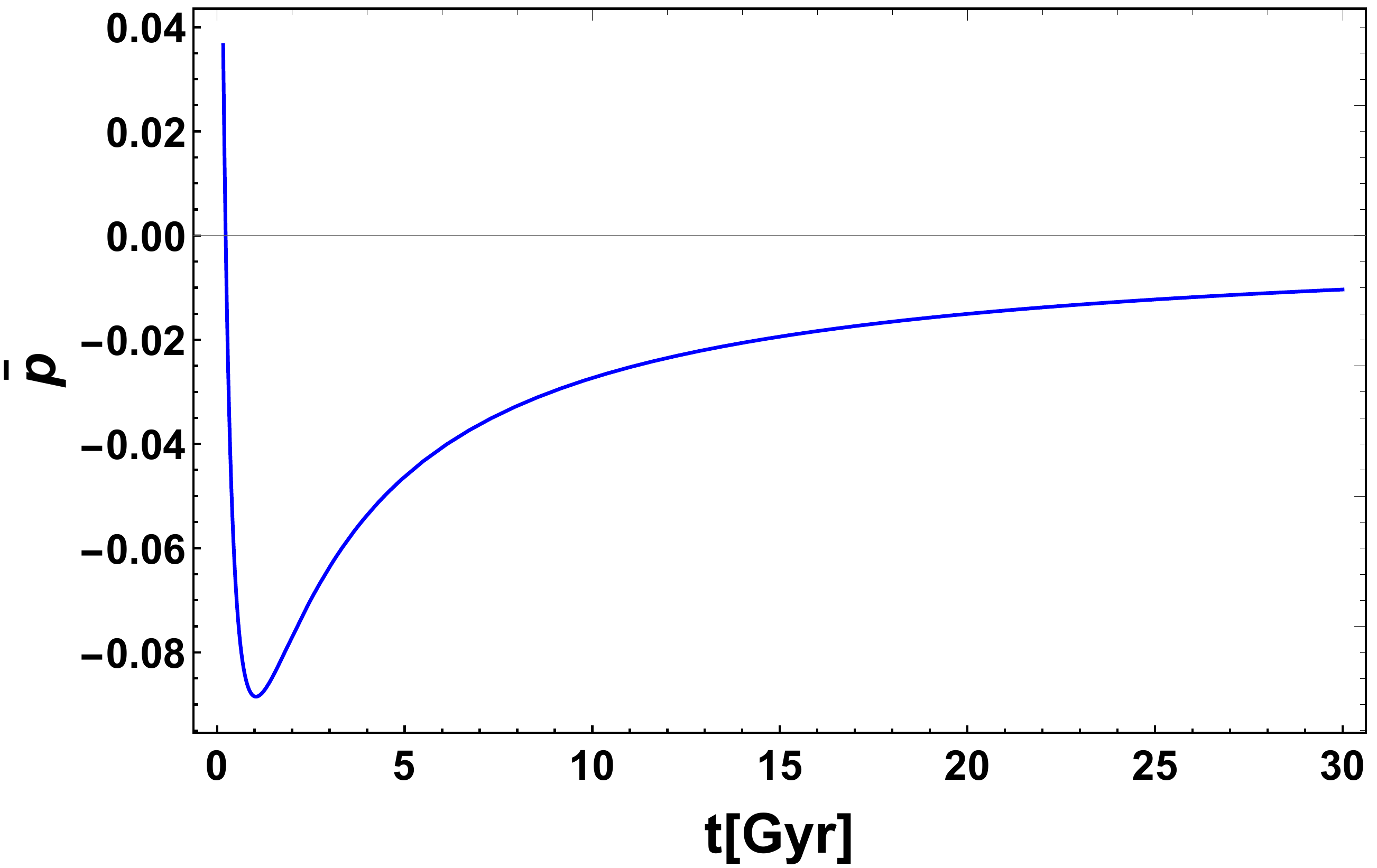}
\caption{Evolution of effective pressure
with $\alpha=-0.179$, $\gamma=0.01$, $\zeta_{0}=-0.0076$, $\zeta_{1}= 0.75$, $\zeta_{2}=-18.4$, $c=14.9$.}
\label{fig6}
\end{figure}

In Fig. \ref{fig5} \& \ref{fig6}, we show the evolution of energy density $\rho$ and effective pressure $\overline{p}$. For an accelerating universe, the pressure has to be negative. Interestingly, within the framework of general relativity,  no known entity posses this feature and therefore to suffice the observations, exotic matter-energy sources must be present. Nonetheless, modified gravity theories provide an alternative route to tackle this enigma by assuming the dark energy to be purely geometrical in nature. The negativity of $\overline{p}$ ensures an accelerating universe at the present epoch. The EoS parameter for our model assumes values close to $-1$ at $t_{0}=13.7$ Gyrs which is in remarkable agreement with the latest Planck measurements \cite{planck}.  

\begin{figure}[H]
\centering
\includegraphics[scale =0.5]{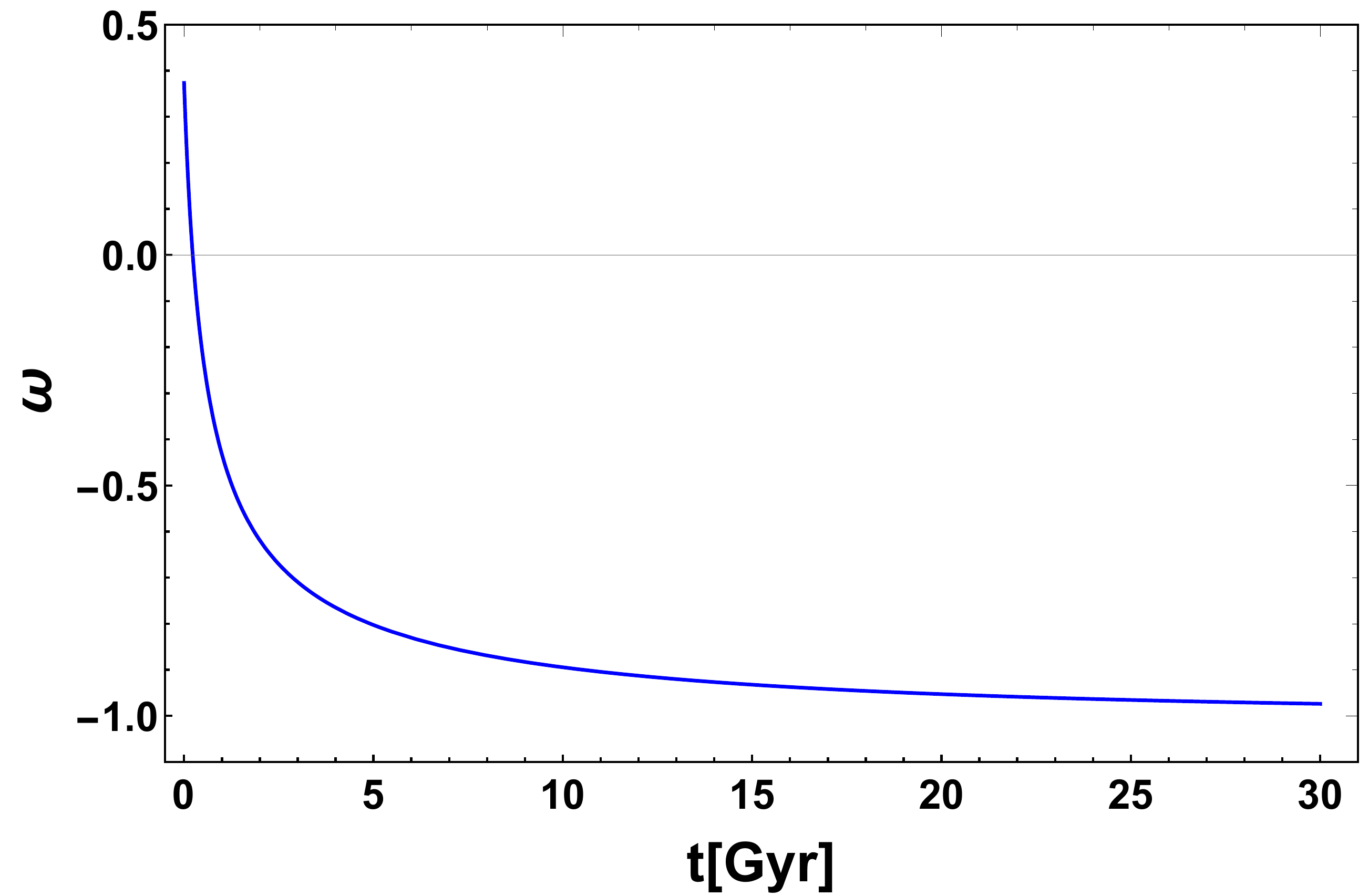}
\caption{Evolution of EoS parameter with $\alpha=-0.179$, $\gamma=0.01$, $\zeta_{0}=-0.0076$, $\zeta_{1}= 0.75$, $\zeta_{2}=-18.4$, $c=14.9$.}
\label{fig7}
\end{figure}

\section{Energy Conditions}\label{V}
The energy conditions are useful linear relationships consisting of energy density and pressure constructed from the Raychaudhuri equation. They are important tools to understand the behavior of lightlike, timelike or spacelike curves and singularities \cite{sahoo,non39} and are defined as:
\begin{itemize}
\item Null energy condition (NEC):  $\rho+p\geq 0$
\item Weak energy conditions (WEC): $\rho\geq 0, \rho+p\geq 0$
\item Strong energy conditions (SEC):$\rho+p\geq 0$, $\rho+3p\geq 0$;
\item Dominant energy conditions (DEC): $\rho\geq 0$, $\rho\geq 0, |p|\leq \rho$. 
\end{itemize}

\begin{figure}[H]
\centering
\includegraphics[scale =0.5]{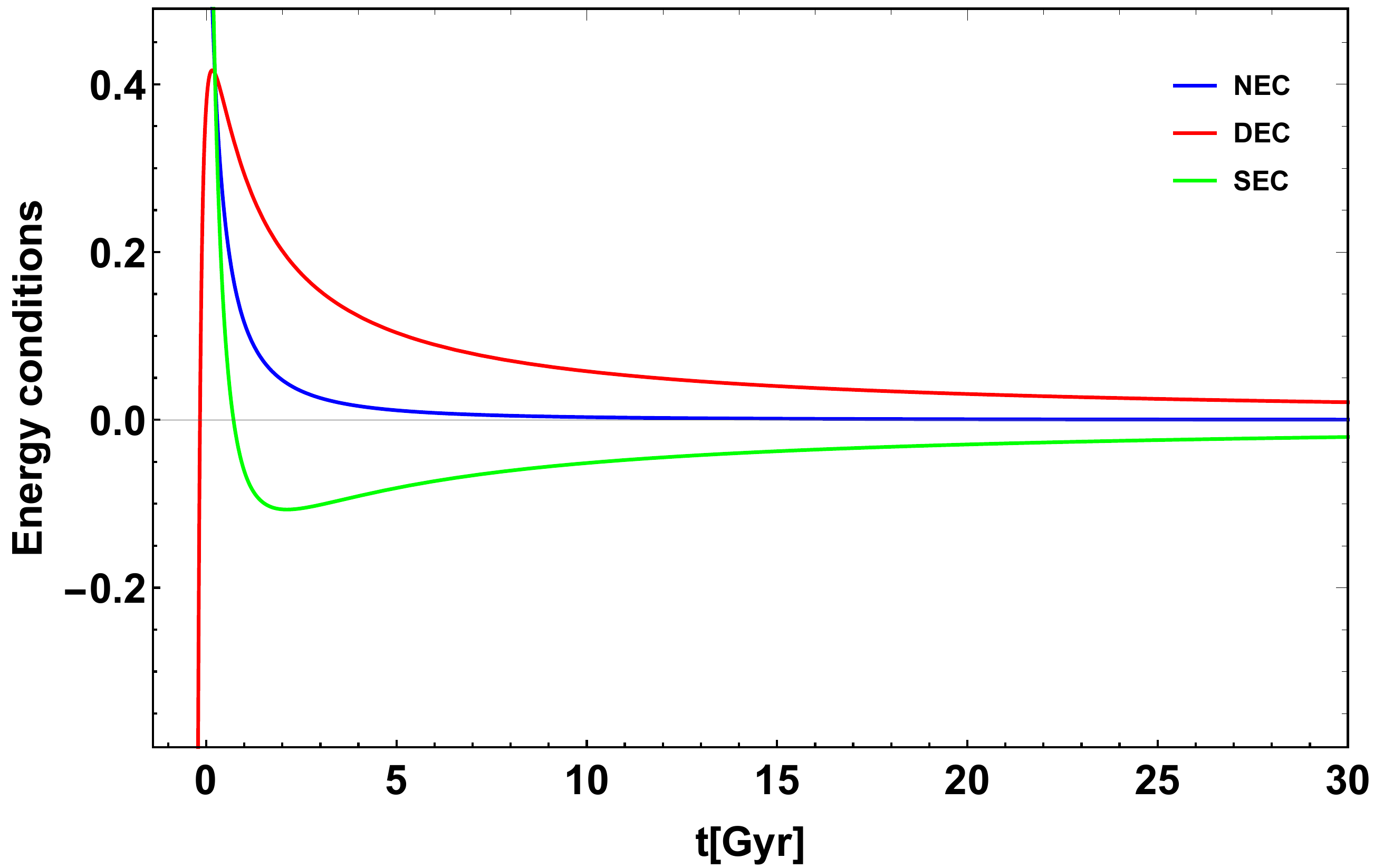}
\caption{Time variation of ECs with $\alpha=-0.179$, $\gamma=0.01$, $\zeta_{0}=-0.0076$, $\zeta_{1}= 0.75$, $\zeta_{2}=-18.4$, c=14.9.}
\label{fig8}
\end{figure}
In Fig. \ref{fig8} we can see the evolution of the enrgy conditions SEC, NEC and WEC as functions of cosmic time. In order to explain the late-time cosmic acceleration with $\omega\simeq-1$, the SEC needs to violate since $p = \omega \rho$. Such a violation of SEC is confirmed from Fig. \ref{fig8} and therefore ensures the cosmological viability of our bulk viscous model.  

 \section{Statefinder Diagnostics}\label{VI}
 
The Statefinder diagnostics is a useful geometrical diagnostic tool capable of distinguishing a wide range of dark energy models from the standard models such as $\Lambda$CDM, HDE, CG, SCDM and Quintessence. The tool consists of two parametric plots: one between  $r$ and $s$ while the other between $r$ and $q$ where $q$ is the deceleration parameter and $r$ and $s$ are defined respectively as \cite{sahni,sahni2}
\begin{equation}
r= \frac{\dddot{a}}{a H^{3}},
\end{equation}

\begin{align*}
s=\frac{r-1}{3\left(q-\frac{1}{2}\right)},\left(q\neq\frac{1}{2}\right).
\end{align*}

The expressions of r and s reads ,
\begin{equation}
\displaystyle r=  -\frac{-\frac{3 (2 \alpha +1) (4 \alpha +1) \zeta_{2} \left(-\frac{9 \left(8 \alpha ^2+6 \alpha +1\right) \zeta_{2}}{\alpha  (\gamma -4)-1}+6 \sqrt{c+\frac{3 \left(8 \alpha ^2+6 \alpha +1\right) \zeta_{2} t}{\alpha  (\gamma -4)-1}}-4 t\right)}{1-\alpha  (\gamma -4)}-4 c}{4 \left(c+\frac{3 \left(8 \alpha ^2+6 \alpha +1\right) \zeta_{2} t}{\alpha  (\gamma -4)-1}\right)}.
\end{equation}

\begin{equation}
\displaystyle s=  -\frac{1-\frac{\left( \frac{3 (2 \alpha +1) (4 \alpha +1) \zeta_{2} \left(-\frac{9 \left(8 \alpha ^2+6 \alpha +1\right) \zeta_{2}}{\alpha  (\gamma -4)-1}+6 \sqrt{c+\frac{3 \left(8 \alpha ^2+6 \alpha +1\right) \zeta_{2} t}{\alpha  (\gamma -4)-1}}-4 t\right)}{1-\alpha  (\gamma -4)}+4 c \right) } {4 \left(c+\frac{3 \left(8 \alpha ^2+6 \alpha +1\right) \zeta_{2} t}{\alpha  (\gamma -4)-1}\right)}}{3 \left(\frac{3 (2 \alpha +1) (4 \alpha +1) \zeta_{2}}{2 (\alpha  (\gamma -4)-1) \sqrt{c+\frac{3 \left(8 \alpha ^2+6 \alpha +1\right) \zeta_{2} t}{\alpha  (\gamma -4)-1}}}-1.5\right)}.
\end{equation}

\begin{figure}[H]
\centering
\includegraphics[scale =0.5]{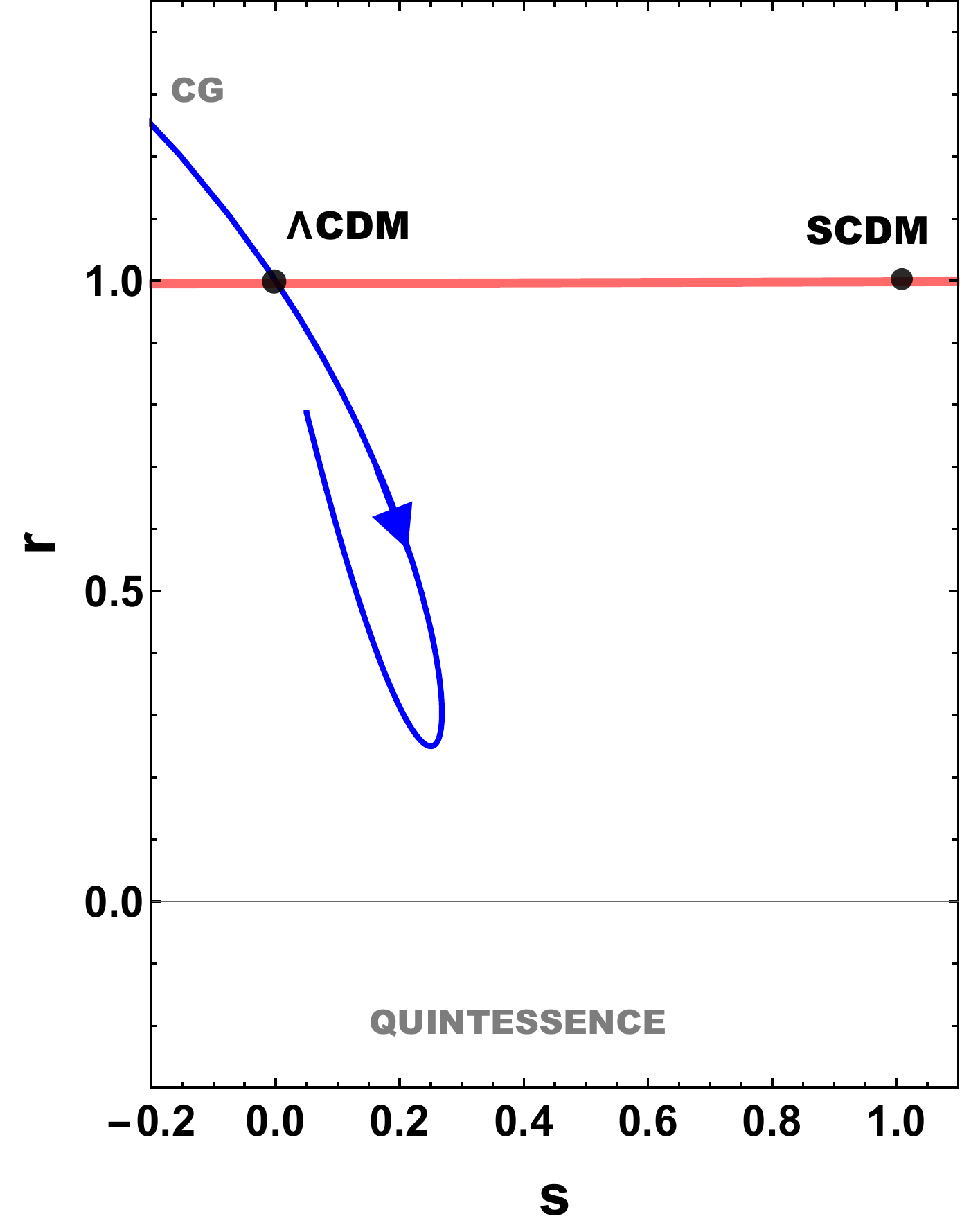}
\caption{$\{r,s\}$ plane for $\alpha=-0.179$, $\gamma=0.01$, $\zeta_{2}=-18.4$, $c=14.9$.}
\label{fig9}
\end{figure}

\begin{figure}[H]
\centering
\includegraphics[scale =0.5]{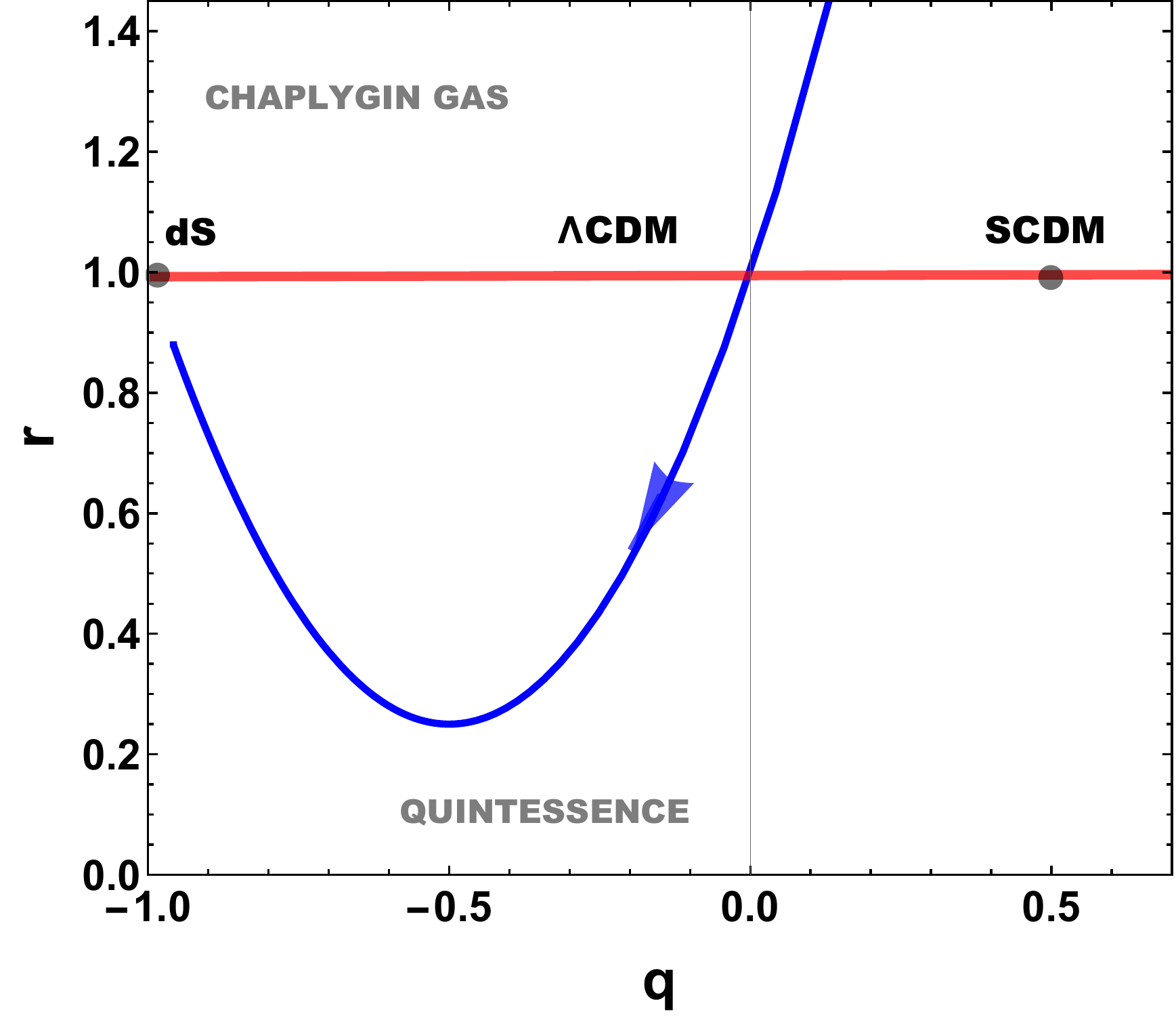}
\caption{$\{r,q\}$ plane for $\alpha=-0.179$, $\gamma=0.01$, $\zeta_{2}=-18.4$, $c=14.9$.}
\label{fig10}
\end{figure}

Different points in this parametric plot corresponds to different dark energy models. In particular,
\begin{itemize}
\item  $\Lambda$CDM corresponds to  $(s=0, r=1)$.
\item HDE corresponds to $(s=\frac{2}{3},r=1)$.
\item CG corresponds to $(s<0, r>1)$.
\item SCDM corresponds to $(s=1, r=1)$.
\item Quintessence corresponds to  $(s>0, r<1)$.
\end{itemize}
The $r-s$ plane is clearly shown in Fig. \ref{fig9} where the arrow indicate the temporal evolution of our model. It can be easily observed that our model behaves like a Chaplygin gas at early times where $(r>1,s<0)$. The model then makes a transition from CG to $\Lambda$CDM and finally stays in the Quintessence region with  $(r<1,s>0)$. In Fig. \ref{fig10} we have shown the $r-q$ plane where the red solid line represent the evolution of $\Lambda$CDM cosmology dividing the plane into two parts. The upper portion belongs to Chaplygin Gas type dark energy models whereas the lower portion corresponding to the Quintessence type dark energy models. The trajectory of our model in the $\{r,q\}$ plane reassures the fact that our model behaves like a CG type dark energy at early times. The model predicts a de-Sitter type expansion with $r=1,q=-1$ in distant future. 

\section{Violation of Energy-Momentum Conservation}\label{VII}

The energy-momentum is conserved in general relativity from the following equation of continuity
\begin{equation}
\dot{\rho}+3H(\rho+p)=0
\end{equation}
implying $d(\rho V)= -p dV $ where $\rho  V$ account for the total energy and $V=a^{3}$ represent the volume of the universe. For a static universe, the total energy remains conserved whereas for an expanding universe, the energy is not conserved and changes with time. Note that according to \cite{harko}, $f(R,T)$ gravity models do not satisfy the law of conservation of energy momentum. In \cite{josset}, the non-conservative cosmological evolution have been investigated by considering dark energy effects as a consequence of energy momentum violation. Also in pioneering studies like \cite{Riess98,Per99}, the cosmic acceleration could itself be a reminiscent of energy-momentum violation on the largest scales. Studies have also been conducted on this subject under $f(R, T)$ gravity \cite{shabani}.\\
Taking covariant derivative of Eq \eqref{e2}, we obtain

\begin{equation}
\nabla^{\mu}T_{\mu\nu}= \frac{f_{T}(R,T)}{1-f_{T}(R,T)} \left[ (T_{\mu\nu}+\Theta_{\mu\nu}) \nabla^{\mu} ln f_{T}(R,T) + \nabla^{\mu} \Theta_{\mu\nu} - \frac{1}{2} g_{\mu\nu}\nabla^{\mu} T \right].
\end{equation}

\begin{figure}[t]
\centering
\includegraphics[scale =0.5]{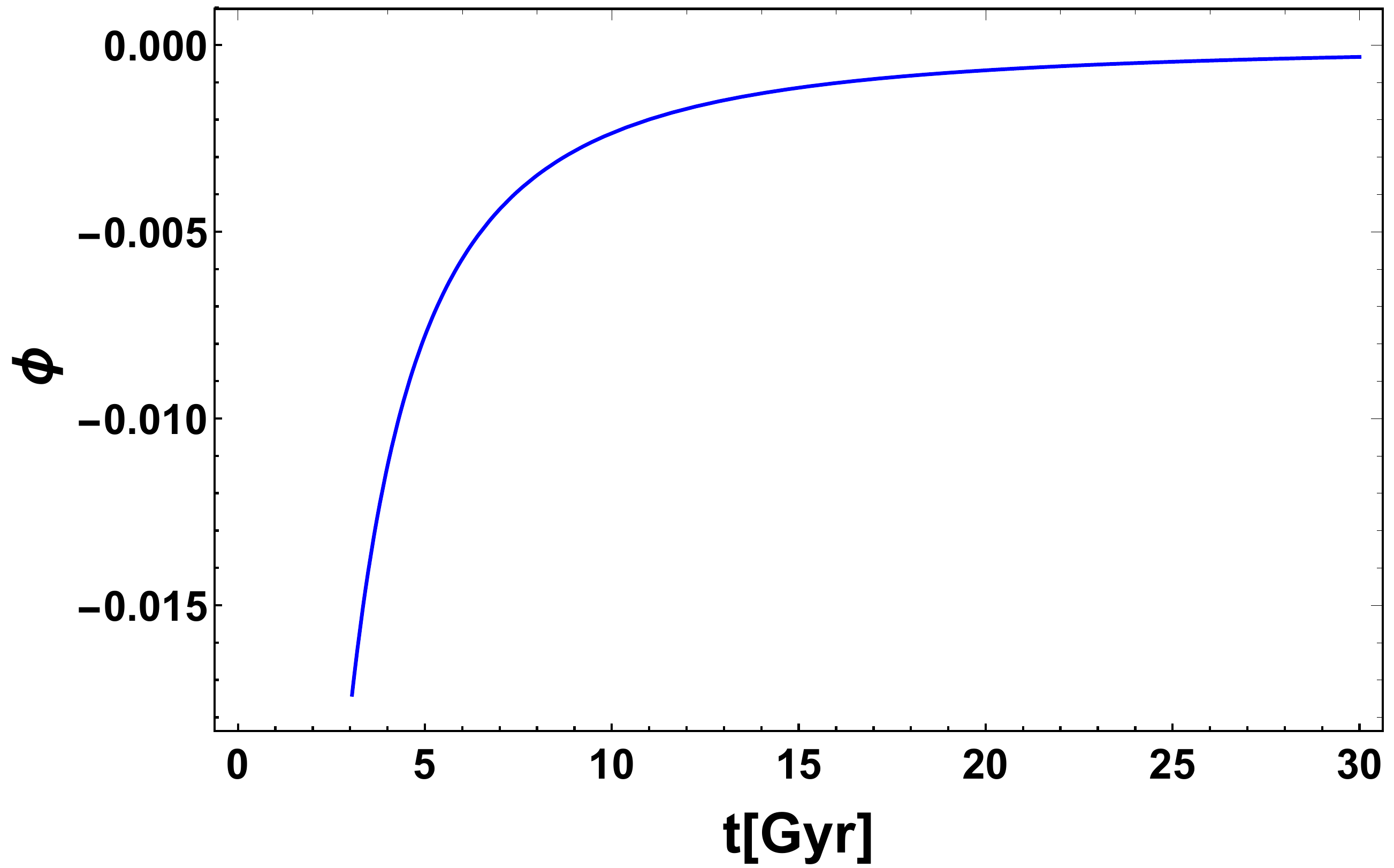}
\caption{Time variation of $\phi$ with $\alpha=-0.179$, $\gamma=0.01$, $\zeta_{0}=-0.0076$, $\zeta_{1}= 0.75$, $\zeta_{2}=-18.4$, c=14.9.}
\label{fig11}
\end{figure}

Substituting $f(R,T)= R+2\alpha T$, yields
\begin{equation}
\nabla^{\mu}T_{\mu\nu}= \frac{-2\alpha}{1+2\alpha} \left[ \nabla^{\mu}(pg_{\mu\nu} +\frac{1}{2} g_{\mu\nu} \nabla^{\mu} T\right]. 
\end{equation}
We note that for $\alpha=0$, $\nabla^{\mu}T_{\mu\nu}=0$ but for $\alpha\neq 0$, there is a violation to the conservation of energy-momentum. Here we estimated the violation of energy-momentum conservation through a deviation factor $\phi$,
\begin{equation}
\phi= \dot{\rho}+3H(\rho+p).
 \end{equation}
in which $\phi=0$ imply conservation of energy-momentum. $\phi$ can vary as positive or negative depending on whether the flow of energy is into the matter field or away from it. From Fig.\ref{fig11} we can observe the non-conservation of energy momentum which decreases with cosmic time.

According to the model the $\phi$ is given as,

\begin{multline}
\displaystyle \phi= \frac{3 \left(\frac{(\gamma -1) \left(-\frac{\alpha  \zeta_{0}}{\sqrt{c+k_{1} t}}+\frac{3 (1-\alpha  \zeta_{1})}{c+k_{1} t}-\frac{\alpha  \zeta _{2}}{(c+k_{1} t)^{3/2}}\right)}{(1+4 \alpha-\alpha \gamma)}+\frac{-\frac{\alpha  \zeta_{0}}{\sqrt{c+k_{1} t}}+\frac{3 (1-\alpha  \zeta_{1})}{c+k_{1} t}-\frac{\alpha  \zeta_{2}}{(c+k_{1} t)^{3/2}}}{(1+4 \alpha-\alpha \gamma)}-\frac{3 \zeta_{0}}{\sqrt{c+k_{1} t}}-\frac{3 \zeta_{1}}{c+k_{1} t}-\frac{3 \zeta_{2}}{(c+k_{1} t)^{3/2}}\right)}{\sqrt{c+k_{1} t}}\\
 +\frac{\frac{\alpha  \zeta_{0} k_{1}}{2 (c+k_{1} t)^{3/2}}-\frac{3 k_{1} (1-\alpha  \zeta_{1})}{(c+k_{1} t)^2}+\frac{3 \alpha  \zeta_{2} k_{1}}{2 (c+k_{1} t)^{5/2}}}{(1+4 \alpha-\alpha \gamma)} . 
\end{multline}

where $k_{1}= -\left( \frac{3(2 \alpha +1)(4 \alpha +1) \zeta_{2}}{\alpha  (-\gamma )+4 \alpha +1}\right)$.

\section{Conclusions}\label{VIII}

In modified gravity theories, dark energy emerges as a result of modified gravitational effects and is purely geometrical in nature. Therefore, a bulk viscous fluid coupled with a modified gravity theory such as $f(R,T)$ gravity could provide an unorthodox way to suffice the cosmic acceleration with just a bulk viscous baryonic fluid.\\
In this paper we studied the phenomena of late-time acceleration by considering a bulk viscous fluid with viscosity coefficient $\zeta =\zeta _{0}+ \zeta _{1} H + \zeta _{2} H^{2}$, where $\zeta _{0}, \zeta _{1}, \zeta _{2}$ are constants in the framework of $f(R,T)$ modified gravity. We solve the field equations with observing the temporal evolution of some kinematic variables such as deceleration parameter, jerk, snap and lerk parameters as a function of cosmic time. We observe the deceleration parameter to be highly sensitive to the model parameter $\alpha$ and undergoes a change in signature at around $t\sim 10$Gyrs for $\alpha=-0.179$ which is favored by observations.  Interestingly, both the jerk and lerk are decreasing function of time whereas the snap is clearly an increasing function. We note that the magnitudes and behaviors of all these parameters strongly suggest an accelerating universe at the present epoch (i.e., $t=13.7$ Gyrs). The EoS parameter for our model assumes values close to $-1$ at $t_{0}=13.7$Gyrs which is in remarkable agreement with the latest Planck measurements \cite{planck}. Our model also show violation of Strong Energy Condition which is required in order to explain the late-time cosmic acceleration with $\omega\simeq-1$, since $p = \omega \rho$. Next, we perform some geometrical diagnostics of the model in $r-s$ and $r-q$ plane. We found that the model is representing a Chaplygin gas type dark energy model at early times while a Quintessence type in distant future. Finally, we  study the violation of continuity equation for our model and show that in order to explain the cosmic acceleration at the present epoch, energy-momentum must violate.

\section*{Acknowledgments}  S.A.
acknowledges CSIR, Govt. of India, New
Delhi, for awarding Junior Research Fellowship. SB thanks Biswajit Pandey for helpful discussions. PKS acknowledges CSIR, New Delhi, India for
financial support to carry out the Research project [No.03(1454)/19/EMR-II
Dt.02/08/2019].

\end{document}